\documentstyle[aps,preprint,epsfig]{revtex}

\begin{document}
\draft \tightenlines

\def\be{\begin{equation}}
\def\ee{\end{equation}}
\def\bea{\begin{eqnarray}}
\def\eea{\end{eqnarray}}

\def\pd{\partial}
\def\a{\alpha}
\def\b{\beta}
\def\g{\gamma}
\def\d{\delta}
\def\m{\mu}
\def\n{\nu}
\def\t{\tau}
\def\l{\lambda}

\preprint{}

\title{Stability and fluctuation modes of giant gravitons with NSNS B field }

\author{ Jin Young Kim\footnote{Electronic address:
jykim@kunsan.ac.kr}}
\address{Department of Physics, Kunsan National University,
Kunsan 573-701, Korea}

\maketitle

\begin{abstract}

 We study the stability of the giant gravitons in the string
 theory background with NSNS B field. We consider the perturbation
 of giant gravitons formed by a probe D$(8-p)$ brane in the
 background generated by D$(p-2)$-D$(p)$ branes
 for $2 \le p \le 5$. We use the quadratic approximation to the
 brane action to find the equations of motion.
 The vibration modes for $\rho$ , $\phi$ and
 $r$ are coupled, while those of $x_k$'s($k= 1,...,p-2)$ are decoupled.
 For $p=5$, they are stable independent of the
 size of the brane. For $p \ne 5$, we calculated the range of the
 size of the brane where they are stable. We also present the mode
 frequencies explicitly for some special cases.

\end{abstract}
\vfill

\newpage

\section{Introduction}

Stable extended brane configurations in some string theory
background, called giant gravitons, attracted interests in
connection with the stringy exclusion principle. Myers
\cite{myers9910053} found that certain D-branes coupled to RR
potentials can expand into higher dimensional branes. McGreevy,
Susskind and Toumbas \cite{mst0003075} have shown that a massless
particle with angular momentum on the ${\rm S}^n$ part of ${\rm
AdS}_m \times {\rm S}^n$ spacetime blows up into a spherical brane
of dimensionality $n-2$. Its radius increases with increasing
angular momentum. The maximum radius of the blown-up brane is
equal to the radius of the sphere that contains it since the
angular momentum is bounded by the radius of ${\rm S}^n$. This is
a realization of the stringy exclusion principle \cite{sep}
through the AdS/CFT correspondence \cite{adscft}. Later it was
shown that the same mechanism can be applied to spherical branes
on the  AdS part \cite{gmt0008015,0008016}.
 However, they can grow arbitrarily large since there
is no upper bound on the angular momentum. To solve this puzzle,
instanton solutions describing the tunneling between the giant
gravitons on the AdS part and on the S part were introduced
\cite{gmt0008015,jlee}. Giant graviton configurations preserving
less than half of the supersymmetry were studied by Mikhailov
\cite{mikhailov}. A magnetic analogue of the Myers effect was
investigated by Das, Trivedi and Vaidya \cite{dtv0008203}. They
suggested that the blowing up of gravitons into branes are
possible on some backgrounds other than ${\rm AdS}_m \times {\rm
S}^n$ spacetime.

Recently it is known by Camino and Ramallo \cite{camram} that the
giant graviton configurations are also possible in a string
background with NS-NS B field. They considered the geometry formed
by a stack of non-threshold bound state of the type (D$(p-2)$,
D$p$) for $2 \le p \le 6$ \cite{bmmcp}, which is characterized by
the non-zero Kalb-Ramond field B from the NS sector together with
the corresponding RR fields. In this background they put a probe
brane such that it could capture both the RR flux and the flux of
B field. The probe brane and the branes of the background have two
common directions. The probe brane is a D$(8-p)$ brane wrapped on
an ${\rm S}^{6-p}$ sphere transverse to the background and
extended along the plane parallel to it. They showed that, for a
particular choice of the worldvolume gauge field, one can find
configurations of the probe brane which behave as massless
particles and they can be interpreted as giant gravitons.

One important issue related to the giant gravitons is whether they
are stable or not under the perturbation around their equilibrium
configurations. Perturbation of the giant gravitons was studied
first by Das, Jevicki and Mathur \cite{djm0009019}. Using the
quadratic approximation to the action, they computed the natural
frequencies of the normal modes for giant gravitons in ${\rm
AdS}_m \times {\rm S}^n$ spacetime for both cases when gravitons
are extended in AdS subspace and they are extended on the sphere
${\rm S}^n$. All modes have real positive $\omega^2$ for any size
of the branes so that they are stable. Perturbation analysis of
giant gravitons whose background geometry is not of a conventional
form of ${\rm AdS}_m \times {\rm S}^n$ was considered by the
author \cite{kimmyung}. The normal modes of giant gravitons in the
dilatonic D-brane background were found and they turned out to be
stable too.

In this paper, we will study the stability analysis and present
the spectrum of the perturbation modes of giant gravitons in the
string background with NS-NS B field described in
Ref.\cite{camram}. We consider the perturbation of giant gravitons
in the near-horizon geometry. In the previous analysis, the
perturbation of the brane along the transverse direction was not
considered. Here we consider the perturbation of this variable
too. The organization of the paper is as follows. In Sec. II we
review the giant gravitons with NSNS B field \cite{camram} and set
up some preliminaries for our calculation. In Sec. III we consider
the perturbation up to second order and derive the equations of
motion from which one determines the normal modes. From these
equations we discuss the stability of the giant gravitons. We also
present the mode frequencies explicitly for some special cases.
Finally in Sec. IV, we conclude and discuss our results.

 \section{Review of giant graviton configuration}

 Consider the supergravity background generated by a stack of $N$
 non-threshold bound states of D$p$ and D$(p-2)$ branes for
 $2 \le p \le 6$. The metric and dilaton are given by
 \cite{camram,bmmcp}

 \be
 ds^2 =  f_p^{-1/2} [ - (dx^0)^2 + \cdots + (d x^{p-2})^2
 + h_p \{ (d x^{p-1})^2 + (d x^p)^2 \} ]
  + f_p^{1/2} ( dr^2 + r^2 d \Omega_{8-p}^2 ) ,
 \label{dsgen}
 \ee

 \be
 e^{ {\tilde \phi}_D } = f_p^{ {3-p} \over 4} h_p^{1/2},
 \label{dilgen}
 \ee
 where $d \Omega_{8-p}^2$is the line element of ${\rm S}^{8-p}$, $r$ is
 the radial coordinate parametrizing  the distance to the brane
 bound state and
 ${\tilde \phi}_D = \phi_D - \phi_D(r  \to \infty )$.
 The functions $f_p$ and
 $h_p$ in Eqs. (\ref{dsgen}) and (\ref{dilgen}) are given by

 \bea
 f_p &=& 1 + { R^{7-p} \over r^{7-p} } , \nonumber  \\
 h_p^{-1} &=& \sin^2 \varphi f_p^{-1} + \cos^2 \varphi , \label{fh}
 \eea
 where $\varphi$ is the angle characterizing the degree of
 mixing of the D$p$ and D$(p-2)$ branes. The parameter $R$ is given by

 \be
 R^{7-p} \cos \varphi = N g_s 2^{5-p} \pi^{{5-p} \over 2}
 ( \alpha^\prime )^{{7-p} \over 2} \Gamma({{5-p} \over 2}) ,
 \label{Rdef}
 \ee
 where $N$ is the number of branes of the stack, $g_s$ is the
 string coupling constant $(g_s = e^{ \phi_D (r  \to \infty )} )$
 and $\alpha^\prime$ is the Regge slope.
 The metric of ${\rm S}^{8-p}$ can be written as

 \be
 d \Omega_{8-p}^2 = {1 \over {1 - \rho^2}} d \rho^2 + (1 - \rho^2) d \phi^2
    + \rho^2 d \Omega_{6-p}^2 ,   \label{domega8mp}
 \ee
 where $d \Omega_{6-p}^2$ is the metric of a unit $6-p$ sphere.
 The range of the variable $\rho$ and $\phi$ are
 $ 0 \le \rho \le 1$ and $ 0 \le \phi \le 2 \pi$ . The coordinate
 $\rho$ plays the role of the size of the system on ${\rm S}^{6-p}$.
 The D$p$ brane
 of the background extends on $x^0 \cdots x^p$, while the D$(p-2)$
 brane lies along $x^0 \cdots x^{p-2}$.
 The component of RR field strengths are:

 \be
 F_{ x^0 , x^1 , \cdots , x^{p-2} , r }^{(p)}
 = \sin \varphi \partial_r f_p^{-1} , ~~~~
 F_{ x^0 , x^1 , \cdots , x^{p} , r }^{(p+2)}
 = \cos \varphi h_p \partial_r f_p^{-1} . \label{RRfield}
 \ee
 The solution also has a rank two NSNS B field on the
 $x^{p-1} x^p$ plane

 \be
 B = \tan \varphi f_p^{-1} h_p d x^{p-1} \wedge dx^p .
 \label{Bfield}
 \ee
 Note that $F^{(p)}$'s for $p \ge 5$ are the hodge duals of those
 with $p \le 5$, i.e. $F^{(p)} =  ^*F^{(10-p)}$. For $\varphi =0 $
 the (D$(p-2)$,D$p$) solution reduces to the D$p$ brane geometry
 whereas for $\varphi = \pi /2 $ it is a D$(p-2)$ brane smeared
 along the $x^{p-1} x^p$ directions.

 Now we embed a probe D$(8-p)$ brane in the near-horizon region of
 the (D$(p-2)$,D$p$) geometry where $r$ is small.
 The probe D$(8-p)$ brane wraps the $(6-p)$ transverse sphere and
 extends along the $x^{p-1} x^{p}$ directions. The action for this
 case, ignoring the fermions, is given by the sum of a
 Dirac-Born-Infeld(DBI) and Wess-Zumino(WZ) terms

 \be
  S = S_{DBI} + S_{WZ} . \label{probeaction}
 \ee
 Taking the worldvolume coordinates
 $\xi^\alpha (\alpha = 0,1, \cdots , 8-p )$ in the static gauge as

 \be
 \xi^\alpha = (t, x^{p-1} ,x^{p} ,\theta^1 ,
     \cdots ,\theta^{6-p} ) , \label{wvcoord}
 \ee
 the dynamical variables are described by

 \be
 r = r(t) , ~~~~\rho = \rho (t) , ~~~~ \phi = \phi (t) .
 \label{rrhophi}
 \ee
 Evaluating the probe brane action under the ansatz of Eq.
 (\ref{rrhophi}), the total action can be written as \cite{camram}

 \be
 S = \int dt d x^{p-1} d x^p {\cal L},
 \ee
 where the lagrangian density ${\cal L}$ is given by

 \bea
 {\cal L} &=&  T_{8-p} \Omega_{6-p} R^{7-p}
   \nonumber  \\
  ~~& \times & \bigg \{ - \rho^{6-p} \lambda_1
 \sqrt{ r^{-2} f_p^{-1} - r^{-2} {\dot r}^{2}
 - { {\dot \rho}^{2} \over {1 - \rho^2} }
 - (1 - \rho^2){\dot \phi}^{2}   }
 + \lambda_2 (-1)^{p+1} \rho^{7-p} {\dot \phi}
 \bigg \} .  \label{lagden}
 \eea
 In Eq. (\ref{lagden}) the functions $\lambda_1$ and $\lambda_2$ are
 defined as

 \be
 \lambda_1 = \sqrt{ h_p f_p^{-1} + {\cal F}^2 h_p^{-1} }, ~~~
 \lambda_2 = F \cos \varphi ,  \label{lambda12}
 \ee
 where $F$ is the only non-zero component of the $U(1)$
 worldvolume  gauge field $F = F_{x^{p-1}, x^p}$, and
 ${\cal F} = F - P[B]$.

 The dynamics of the system is determined by the standard
 hamiltonian analysis. Absorbing the $(-1)^{p+1}$
 sign into the redefinition of ${\dot \phi}$ if
 necessary, the conjugate momenta are calculated as

 \be
 {\cal P}_q = { { \partial {\cal L} } \over { \partial {\dot q} } }
  \equiv T_{8-p} \Omega_{6-p} R^{7-p} \lambda_1 \pi_q ,
 \label{conjmom}
 \ee
 for $q = r,~ \rho $ and $\phi$, where $\pi_q$'s are defined as
 \bea
 \pi_r &=& {\rho^{6-p} \over r^2}
 {  {\dot r} \over
 \sqrt{ r^{-2} f_p^{-1} - r^{-2} {\dot r}^{2}
 - { {\dot \rho}^{2} \over {1 - \rho^2} }
 - (1 - \rho^2){\dot \phi}^{2}   }   } , \nonumber  \\
 \pi_\rho &=& {\rho^{6-p} \over {1-\rho^2} }
 {  {\dot \rho} \over
 \sqrt{ r^{-2} f_p^{-1} - r^{-2} {\dot r}^{2}
 - { {\dot \rho}^{2} \over {1 - \rho^2} }
 - (1 - \rho^2){\dot \phi}^{2}   }   } , \nonumber  \\
 \pi_\phi &=& ( 1- \rho^2) \rho^{6-p}
 {  {\dot \phi} \over
 \sqrt{ r^{-2} f_p^{-1} - r^{-2} {\dot r}^{2}
 - { {\dot \rho}^{2} \over {1 - \rho^2} }
 - (1 - \rho^2){\dot \phi}^{2}   }   }
 + \Lambda \rho^{7-p} . \label{reducedmom}
 \eea
 In the third expression of Eq. (\ref{reducedmom}),
 $\Lambda$ is defined as
 $ \Lambda =  \lambda_1 / \lambda_2 $.
 The hamiltonian density can be calculated as

 \be
 {\cal H} = {\dot r} {\cal P}_r + {\dot \rho} {\cal P}_{\rho}
  + {\dot \phi} {\cal P}_\phi - {\cal L}
 \equiv  T_{8-p} \Omega_{6-p} R^{7-p} \lambda_1 h ,
 \ee
 where $h$ the reduced hamiltonian in analogy with
 the reduced ones for momenta

 \be
 h = r^{-1} f_p^{-{1 \over 2}} \Bigg [
  r^2 \pi_r^2 + \rho^{2(6-p)} + (1- \rho^2) \pi_\rho^2
  + { {(\pi_\phi - \Lambda \rho^{7-p} )^2} \over {1 - \rho^2} }
  \Bigg ]^{1 \over 2} . \label{reducedh}
 \ee

 We consider the solution of the equations of motion
 derived from the reduced hamiltonian Eq. (\ref{reducedh}). From Eq.
 (\ref{domega8mp}) the coordinate $\rho$ plays the role of the
 size of the system on $S^{6-p}$ shpere. For this reason we look
 for the solution of the equations of motion with constant $\rho$
 which corresponds to the giant graviton configuration.
 The same
 problem was considered in Ref. \cite{dtv0008203} for the case of
 probe branes moving  in the near-horizon D$p$ brane background.
 Comparing the right-hand side of Eq. (\ref{reducedh}) with the
 corresponding expression in Ref. \cite{dtv0008203}, the same kind
 of arrangement is possible if the condition

 \be
 \Lambda = {\lambda_1 \over \lambda_2} = 1 , \label{Lambdae1}
 \ee
 is satisfied. Indeed, if this condition is satisfied, $h$ can be
 expressed as

 \be
 h = r^{-1} f_p^{-{1 \over 2}} \Bigg [
  \pi_\phi^2 + r^2 \pi_r^2 + (1- \rho^2) \pi_\rho^2
  + { {(\pi_\phi \rho - \rho^{6-p} )^2} \over {1 - \rho^2} }
  \Bigg ]^{1 \over 2} . \label{reducedh2}
 \ee
 We can find the brane configuration with constant $\rho$ under the
 condition Eq. (\ref{Lambdae1}). From Eq. (\ref{reducedmom}), we have
 $ \pi_\rho = 0$.
 Then, from the hamiltonian equation of motion for $\pi_\rho$ ,
 i.e. ${\dot \pi}_\rho = - \partial h / \partial \rho$, the last
 term on the right-hand side of Eq. (\ref{reducedh2}) must vanish.
 For $p< 6$ this happens either for

 \be
 \rho =0,    \label{rhoezero}
 \ee
 or when $\pi_\phi$ is given by

 \be
 \pi_\phi = \rho^{5-p} . \label{piphi}
 \ee
 For $p = 6$, only Eq. (\ref{piphi}) gives the constant $\rho$
 configuration. Since
 $h$ does not depend on $\phi$ explicitly, $\pi_\phi$ is a constant
 of motion. Thus, for $p \ne 5$, Eq. (\ref{piphi}) makes sense
 only when $\rho$ is constant. Actually the constant value of
 $\rho$ is determined by the value of $\pi_\phi$. When $p=5$,
 $\pi_\phi = 1$ regardless of the value of $\rho$.

 Taking $\Lambda =1$, from the last expression of Eq.
 (\ref{reducedmom}), ${\dot \phi}$ is calculated as

 \be
  {\dot \phi} =
  {  { \pi_\phi - \rho^{7-p} } \over  { 1 - \rho^2} }
  { { \bigg [ r^{-2} ( f_p^{-1} - {\dot r}^{2} )
   - { {\dot \rho}^{2} \over {1 - \rho^2} } \bigg ]^{1 \over 2} }
 \over
 { \bigg [ \pi_\phi^2 +
    { {( \pi_\phi \rho - \rho^{6-p} )^2 } \over {1 - \rho^2} }
   \bigg ]^{1 \over 2} } }. \label{phidotexp}
 \ee
 Since ${\dot \rho} =0$, one can easily check that
 ${\dot \phi}$ and ${\dot r}$ satisfy

 \be
 f_p (r^2 {\dot \phi}^2 + {\dot r}^2 ) = 1,  \label{velrel}
 \ee
 whenever one of the two conditions in Eq. (\ref{rhoezero})
 or Eq. (\ref{piphi})
 is met. For the configurations we are considering, the last two
 terms inside the square root of the reduced hamiltonian
 Eq. (\ref{reducedh2}) vanish
 and this configuration certainly minimizes the energy.
 Eq. (\ref{velrel}) is the condition satisfied by a particle moving in
 the $(r, \phi)$ plane at $\rho =0$ along a null trajectory in the
 metric Eq. (\ref{dsgen}). Thus the configurations have the
 characteristic of a massless particle i.e. the giant graviton.

 The momentum
 $p_\phi$ and $p_r$ can be obtained by integrating the
 momentum densities ${\cal P}_\phi$
 and ${\cal P}_r$ over the $x^{p-1} x^p$ plane.
 The momentum density ${\cal P}_\phi$ can be obtained from Eqs.
  (\ref{conjmom}) and (\ref{piphi})

 \be
 {\cal P}_\phi = {T_f \over {2 \pi} } F N \rho^{5-p} .
 \label{calpphiform}
 \ee
 and $p_\phi$ is obtained as

 \be
 p_\phi = \int dx^{p-1} dx^p {\cal P}_\phi
 = N N' \rho^{5-p},
 \ee
 where $N'$ is the total flux defined by

 \be
  {T_f \over {2 \pi} } \int d x^{p-1} d x^p F =  N'
 \ee
 For $p <5$, the size of the wrapped brane increases with the
 momentum $p_\phi$. Since $0 \le \rho \le 1$, the maximum value of
 the momentum is

 \be
 p_\phi^{max} = N N'     \label{pphimax}
 \ee
 for $\rho =1$.
 It is known that the existence of the maximum angular momentum
 is the manifestation of the stringy
 exclusion principle \cite{sep}. For $p=5$ the momentum $ p_\phi $ is
 independent of the value of $\rho$. For $p=6$, the value in
 Eq. (\ref{pphimax}) is actually the minimum. We will not consider
 $p=6$ case significantly since we cannot
 define angular momentum on ${\rm S}^{6-p}$ for $p=6$.

 The energy of the giant graviton can be obtained by
 integrating the hamiltonian density ${\cal H}$ over the $x^{p-1} x^p$ plane

 \be
 H_{GG} = f_p^{- {1 \over 2} }
 \bigg ( p_r^2 + { {p_\phi^2} \over r^2 } \bigg )^{1 \over 2}
 = R^{ {p-7} \over 2 }
 \bigg ( r^{7-p} p_r^2 + r^{5-p} {p_\phi^2} \bigg )^{1 \over 2} .
 \label{HGG}
 \ee
 Requiring the conservation of energy $H_{GG} = E$,
 we have

 \be
 {\dot r}^2 + { r^{7-p} \over R^{7-p} }
 \Bigg ( { {p_\phi^2} \over {E^2 R^{7-p} } } r^{5-p} - 1
 \Bigg ) = 0 .
 \label{rdoteq}
 \ee
 The solution of this differential equation, for $p \ne 5$, is
 given by

 \be
  \bigg ({r_* \over r} \bigg )^{5-p} = 1 + (5-p)^2
  r_*^{5-p} R^{p-7}  \bigg ({{t -t_*} \over 2} \bigg )^2,
  ~~~~~~~~~(p \ne 5),
 \ee
 where $r_*$ is defined as

 \be
 (r_*)^{5-p} = {E^2 \over p_\phi^2} R^{7-p}.
 \ee
 Note that $t_*$ is the value at which $r = r_*$. So we can take $t_*=0$
 without loss of generality. For $p <5$,
 $r \to 0$ as $t \to \pm \infty$, i.e. the giant graviton
 falls asymptotically to the horizon. However, for $p=6$,
 $r \to \infty$ as $t \to \pm \infty$, i.e. it always escapes
 to infinity.
 The solution for $p=5$ is a special case and easier to integrate

 \be
 r = r_0 e^{ \pm {t \over R}
 \sqrt{ 1 -{ {p_\phi^2} \over {E^2 R^2 } } } },
 ~~~~~~~~~(p = 5) .
 \ee
 The solution connects asymptotically the point $r=0$ and
 $r = \infty$.

 Similarly we can express $\phi$ as a function of $t$.
 We substitute the $r(t)$ expression into Eq.
 (\ref{velrel}) to get ${\dot \phi}$ then we integrate it over $t$.
 The result for $p \ne 5$ is

 \be
 \tan \Bigg [ { {5-p} \over 2} ( \phi - \phi_*) \Bigg ]
 ={ {5-p} \over 2} \Bigg ( {r_* \over R} \Bigg )^{ {5-p} \over 2}
 {t \over R} , ~~~~~~~~~( p \ne 5) ,
 \ee
 and for $p =5$,

 \be
 \phi = \phi_* + {p_\phi \over {E R^2} } t , ~~~~~~~~~(p = 5) .
 \ee

 \section{Stability analysis and mode frequencies }

 In the previous section we have reviewed how the giant graviton
 picture appears in the near horizon background of D$(p-2)$-D$p$
 branes. Here we will consider the perturbations of the giant
 gravitons from the equilibrium configurations. A small vibration of
 the brane can be described by defining spacetime coordinates
 $(r , \rho ,\phi , x_k (k=1, \cdots , p-2) )$ as functions of the
 worldvolume coordinates $(t, x^{p-1} , x^p , \theta^1 , \cdots ,
 \theta^{6-p} )$

 \bea
 r &=& r_0 (t) + \epsilon~ \delta r (t, x^{p-1} , x^p , \theta^1 , \cdots ,
    \theta^{6-p} ) ,   \nonumber  \\
 \rho &=& \rho_0 + \epsilon~ \delta \rho (t, x^{p-1} , x^p , \theta^1 ,
    \cdots , \theta^{6-p} ) ,   \nonumber   \\
 \phi &=& \phi_0 (t) + \epsilon~ \delta \phi (t, x^{p-1} , x^p , \theta^1 ,
 \cdots , \theta^{6-p} ) ,   \nonumber   \\
 x_k  &=&  \epsilon~ \delta x_k (t, x^{p-1} , x^p , \theta^1 ,
 \cdots , \theta^{6-p} ) , ~~~~~~(k=1, \cdots , p-2) .
 \label{pertvarform}
 \eea
 Here $\rho_0$ is a constant with $\rho_0 =0$ or $\rho_0^{5-p} =
 \pi_\phi$. $r_0 (t)$ and $\phi_0 (t)$ are the solutions of the
 unperturbed equilibrium configuration found in the previous
 section. The action of the probe brane can be expanded in
 orders of $\epsilon$ as

 \be
 S = \int dt dx^{p-1} dx^p d\theta^1 \cdots d\theta^{6-p}
  \bigg \{ {\cal L}_0 + {\cal L}_1 (\epsilon)
  + {\cal L}_2 (\epsilon^2)+ \cdots \bigg \} .
 \ee
 Obviously ${\cal L}_0$ gives the zeroth order lagrangian
 density that we have used in Sec. II.

 \subsection{ First order perturbation}

 First we expand the action to linear order in $\epsilon$. We
 substitute Eq. (\ref{pertvarform}) into the brane action Eq.
 (\ref{probeaction}) and a straightforward calculation gives

 \bea
 {\cal L}_1 (\epsilon) &=& \epsilon T_{8-p}  R^{7-p}
   \sqrt{ {\hat g}^{6-p} } \lambda_1  \nonumber   \\
 &\times& \Bigg [  \delta \rho \bigg \{ -
 {  { \rho_0^{7-p} {\dot \phi}_0^2 } \over
     \sqrt{ { r_0^{5-p} \over R^{7-p} } - { {\dot r}_0^2 \over r_0^2 } -
        (1 - \rho_0^2) {\dot \phi}_0^2 }   }  \nonumber  \\
 &&~~~~~~~-(6-p)\rho_0^{5-p}
  \sqrt{ { r_0^{5-p} \over R^{7-p} } - { {\dot r}_0^2 \over r_0^2 } -
        (1 - \rho_0^2) {\dot \phi}_0^2 }
 + (7-p) \rho_0^{6-p} {\dot \phi}_0  \bigg \} \nonumber   \\
 &+& {\dot {\delta \phi} } \bigg \{
 {  { (1- \rho_0^2) \rho_0^{6-p} {\dot \phi}_0 } \over
  \sqrt{ { r_0^{5-p} \over R^{7-p} } - { {\dot r}_0^2 \over r_0^2 } -
        (1 - \rho_0^2) {\dot \phi}_0^2 }  }
     + \rho_0^{7-p} \bigg \}    \nonumber   \\
 &+& { {\delta r}  \over r_0 } \bigg \{
  - {  { \rho_0^{6-p} } \over
 \sqrt{ { r_0^{5-p} \over R^{7-p} } - { {\dot r}_0^2 \over r_0^2 } -
        (1 - \rho_0^2) {\dot \phi}_0^2 }  }
 \bigg( { {5-p} \over 2} { r_0^{5-p} \over R^{7-p} }
           + { {\dot r}_0^2  \over r_0^2 }
  \bigg )      \bigg \}    \nonumber   \\
 &+& { {\dot {\delta r} } \over {\dot r_0} } \bigg \{
   {  { \rho_0^{6-p} } \over
 \sqrt{ { r_0^{5-p} \over R^{7-p} } - { {\dot r}_0^2 \over r_0^2 } -
        (1 - \rho_0^2) {\dot \phi}_0^2 }   }
            { {\dot r}_0^2  \over r_0^2 }     \bigg \}
 \Bigg ] .   \label{firstordact}
 \eea
 If we substitute

 \be
 {\dot \phi}_0^2 = r_0^{-2} ( f_p^{-1} - {\dot r}_0^2 ) ,
 \label{velvel2}
 \ee
  obtained from Eq. (\ref{velrel}),
 the square root in Eq. (\ref{firstordact}) is just
 $\rho_0 {\dot \phi}_0$. Then one can
 easily check that the coefficient of $\delta \rho$ vanishes. The
 coefficient of ${\dot {\delta \phi} }$ is constant
 ($\rho_0^{5-p}$) and thus this term does not contribute to the
 variation of the action with fixed boundary values.
 The last term (${\dot {\delta r} }$ term) can be
 integrated by parts with respect to $t$. Neglecting the
 boundary terms, this term can be replaced by

 \be
 \int dt {\dot {\delta r} } \bigg (
 { {\rho_0^{5-p} {\dot r}_0} \over { {\dot \phi}_0 r_0^2 } }
 \bigg )
 = - \int d t {\delta r} {d \over dt} \bigg (
 { {\rho_0^{5-p} {\dot r}_0} \over { {\dot \phi}_0 r_0^2 } }
 \bigg ).
 \ee
 Combining this term with the third term(${\delta r}$ term),
 the coefficient of $\delta r$ is

 \bea
 &-& \rho_0^{5-p} \Bigg [ {1 \over {r_0 {\dot \phi} } }
  \bigg ( { { 5-p} \over 2}
 { r_0^{5-p} \over R^{7-p} }
           + { {\dot r}_0^2  \over r_0^2 }   \bigg )
 + {d \over dt} \bigg (
 { {\dot r}_0  \over { {\dot \phi}_0 r_0^2 } }
 \bigg )       \Bigg ]   \nonumber   \\
 = &-& \rho_0^{5-p}  {1 \over {r_0 {\dot \phi}_0 } }
  \bigg ( { { 5-p} \over 2} { r_0^{5-p} \over R^{7-p} }
  - { {\dot r}_0^2  \over r_0^2 }
 -{ {\dot r}_0  \over r_0 } { {\ddot \phi}_0  \over {\dot \phi}_0 }
 + { {\ddot r}_0  \over r_0 }       \bigg ) .
 \label{deltarterm2}
 \eea
 To simplify this expression we substitute the following equation,

 \be
  {\ddot \phi}_0
  = { 1 \over {\dot \phi}_0 } \bigg (
  { {5-p} \over 2 } { r_0^{4-p} \over R^{7-p} } {\dot r}_0
 + { {\dot r}_0^3  \over r_0^3 }
 -  { { {\dot r}_0  {\ddot r}_0 } \over r_0^2 }  \bigg ) ,
 \label{dddotphi}
 \ee
 obtained from Eq. (\ref{velvel2}), into Eq. (\ref{deltarterm2}), the
 coefficient of $\delta r$ is calculated as

 \bea
  &-& \rho_0^{5-p}  {1 \over {r_0 {\dot \phi}_0^3 } }
 \Bigg [
  \bigg ( { { 5-p} \over 2} { r_0^{5-p} \over R^{7-p} }
  - { {\dot r}_0^2  \over r_0^2 } + { {\ddot r}_0 \over r_0 } \bigg )
  {\dot \phi}_0^2
  - \bigg (
  { {5-p} \over 2 } { r_0^{3-p} \over R^{7-p} } {\dot r}_0^2
 + { {\dot r}_0^4  \over r_0^4 }
 -  { { {\dot r}_0^2  {\ddot r}_0 } \over r_0^3 }  \bigg )
  \Bigg ]               \nonumber     \\
  = &-& \rho_0^{5-p}  {1 \over {r_0 {\dot \phi}_0^3 } }
 \Bigg [
   { { 5-p} \over 2} { r_0^{2(5-p)} \over R^{2(7-p)} }
  - (6-p) { r_0^{3-p} \over R^{7-p} } {\dot r}_0^2
  - { r_0^{4-p} \over R^{7-p} } {\ddot r}_0
  \Bigg ] .      \label{deltarterm3}
 \eea
 We have used Eq. (\ref{velvel2}) in the the above
 equation to get the second line.
 This expression can be simplified further. Differentiating Eq.
(\ref{rdoteq}), we have
 \be
 {\ddot r}_0 = -(6-p) { {p_\phi^2} \over {E^2 R^{2(7-p)} } }
 r_0^{11-2p} + { {7-p} \over 2} { r_0^{6-p} \over R^{7-p} } .
 \label{rddoteq}
 \ee
 Substituting Eqs. (\ref{rdoteq}) and (\ref{rddoteq}) into
 Eq. (\ref{deltarterm3}), one can show that the square bracket is
 just zero. Thus, we find that the first order term in
 $\epsilon$ vanishes. This confirms that the zeroth order solution
 described in Sec. II is the right solution which minimizes the action.

 \subsection{ Second order perturbation}

 Now we consider the second order term in $\epsilon$. The second
 order term is calculated as

\bea
 {\cal L}_2 (\epsilon^2) &=& - {\epsilon^2 \over 2} T_{8-p} R^{7-p}
  \rho_0^{7-p} \lambda_1
  \omega_0 \sqrt{ {\hat g}^{6-p} }     \nonumber   \\
 \times  \Bigg [
  &-& { 1 \over { \rho_0^2 (1 - \rho^2) \omega_0^2} }
  ( {\dot {\delta \rho} })^2
  + \sum_{i=1}^{6-p} { 1 \over { \rho_0^2 (1 - \rho^2)} }
 \bigg ( { {\partial \delta \rho} \over {\partial \theta_i} }
       \bigg )^2 {\hat g}^{\theta_i \theta_i}
 + {1 \over \lambda_1^2} \sum_{j=p-1}^{p}
 { r_0^2 \over { 1 - \rho^2} }
 \bigg ( { {\partial \delta \rho} \over {\partial x^i} }
       \bigg )^2         \nonumber    \\
  &-& { {1 - \rho^2} \over { \rho_0^4  \omega_0^2} }
  ( {\dot {\delta \phi} })^2
  + \sum_{i=1}^{6-p} { {1 - \rho_0^2} \over  \rho_0^2  }
 \bigg ( { {\partial \delta \phi} \over {\partial \theta_i} }
       \bigg )^2 {\hat g}^{\theta_i \theta_i}
 + {1 \over \lambda_1^2} \sum_{j=p-1}^{p}
  r_0^2 ( 1 - \rho_0^2)
 \bigg ( { {\partial \delta \phi} \over {\partial x^i} }
       \bigg )^2         \nonumber    \\
 &-& {1 \over { \rho_0^2 r_0^2  \omega_0^2} }
  ( {\dot {\delta r} })^2
  + \sum_{i=1}^{6-p} { 1 \over {r_0^2 \rho_0^2 } }
 \bigg ( { {\partial \delta r} \over {\partial \theta_i} }
       \bigg )^2 {\hat g}^{\theta_i \theta_i}
 + {1 \over \lambda_1^2} \sum_{j=p-1}^{p}
  \bigg ( { {\partial \delta r} \over {\partial x^j} }
       \bigg )^2         \nonumber    \\
  &&~~~~~~~~~~~~~~~ + {1 \over { \rho_0^2 r_0^2 } }
  {{5-p} \over 2} \bigg ( 4-p - {{5-p} \over 2}
  {1 \over { \rho_0^2 } } \bigg )
  ( {\delta r} )^2     \nonumber    \\
 &+& \sum_{k=1}^{p-2}  \bigg \{
 - {1 \over \rho_0^2 } ( {\dot {\delta x_k} })^2
  + \sum_{i=1}^{6-p} { \omega_0^2 \over \rho_0^2  }
 \bigg ( { {\partial \delta x_k} \over {\partial \theta_i} }
       \bigg )^2 {\hat g}^{\theta_i \theta_i}
 + {1 \over \lambda_1^2} \sum_{j=p-1}^{p} \omega_0^2 r_0^2
  \bigg ( { {\partial \delta x_k } \over {\partial x^i} }
       \bigg )^2   \bigg \}  \nonumber   \\
 &+& (5-p) \bigg (
   - {2 \over {\rho_0^3 \omega_0 } }
  {\delta \rho} {\dot {\delta \phi} }
  + {  {5-p} \over {\rho_0^3 r_0 } }
  {\delta \rho} {\delta r}
  + {{1 - \rho_0^2} \over {\rho_0^4 \omega_0 r_0} }
  {\dot {\delta \phi} } {\delta r}
                \bigg )
                \Bigg ],
 \eea
 where $\omega_0 = {\dot \phi}$.
 In general $\omega_0$ and ${\dot r}_0$ in the above expression are
 functions of time, so we have to consider the time dependence of
 these variables in deriving the equations of motion. However,
 since our formalism is valid only for the near-horizon region
 ($r << R$), we can consider the perturbation around this region.
 Thus we treat $\omega_0$ and $r_0$ as time independent
 values around the equilibrium configuration at $t=t^* =0$.
 The equations of motion for this case are given by

 \bea
   { 1 \over { \rho_0^2 (1 - \rho_0^2) \omega_0^2} }
   {\ddot {\delta \rho} }
  &-& { 1 \over { \rho_0^2 (1 - \rho_0^2)} } \sum_{i=1}^{6-p}
 {\partial \over {\partial \theta_i} }
 \bigg ( { {\partial \delta \rho} \over {\partial \theta_i} }
        {\hat g}^{\theta_i \theta_i}  \bigg )
 - {1 \over \lambda_1^2} { r_0^2 \over { 1 - \rho_0^2} }
 \sum_{j=p-1}^{p} {\partial \over {\partial x^j} }
 \bigg ( { {\partial \delta \rho} \over {\partial x^j} }
       \bigg )  \nonumber   \\
  &-& (5-p) { 1 \over { \rho_0^3 \omega_0} } {\dot {\delta \phi} }
 + { {(5-p)^2} \over 2 } { 1 \over { \rho_0^3 r_0} } {\delta r }
  = 0 , \label{delrhoe}
 \eea

 \bea
   { {1 - \rho_0^2} \over { \rho_0^4  \omega_0^2} }
   {\ddot {\delta \phi} }
  &-& { {1 - \rho_0^2} \over  \rho_0^2 } \sum_{i=1}^{6-p}
 {\partial \over {\partial \theta_i} }
 \bigg ( { {\partial \delta \phi} \over {\partial \theta_i} }
        {\hat g}^{\theta_i \theta_i}  \bigg )
 - {1 \over \lambda_1^2} r_0^2 ( 1 - \rho_0^2 )
 \sum_{j=p-1}^{p} {\partial \over {\partial x^j} }
 \bigg ( { {\partial \delta \phi} \over {\partial x^j} }
       \bigg )  \nonumber   \\
  &+& (5-p) { 1 \over { \rho_0^3 \omega_0} } {\dot {\delta \rho} }
 - { {5-p} \over 2 } { {1 - \rho_0^2} \over { \rho_0^4 \omega_0 r_0} }
 {\dot {\delta r } }  = 0 ,   \label{delphie}
 \eea

 \bea
 &&{ 1 \over { \rho_0^2 r_0^2 \omega_0^2} } {\ddot {\delta r} }
  - { 1 \over { \rho_0^2 r_0^2} } \sum_{i=1}^{6-p}
 {\partial \over {\partial \theta_i} }
 \bigg ( { {\partial \delta r} \over {\partial \theta_i} }
        {\hat g}^{\theta_i \theta_i}  \bigg )
 - {1 \over \lambda_1^2}
 \sum_{j=p-1}^{p} {\partial \over {\partial x^j} }
 \bigg ( { {\partial \delta r} \over {\partial x^j} }
       \bigg )  \nonumber   \\
  &+& { {5-p} \over 2 } { 1 \over { \rho_0^2 r_0^2} }
 \bigg ( 4-p - { {5-p} \over 2 } {1 \over \rho_0^2}  \bigg )
   {\delta r }
 + { {(5-p)^2} \over 2 } { 1 \over { \rho_0^3 r_0} } {\delta \rho }
 + { {5-p} \over 2 } { {1 - \rho_0^2} \over { \rho_0^4 r_0 \omega_0} }
 {\dot {\delta \phi} }
  = 0 ,   \label{delre}
 \eea

 \be
 { 1 \over  \rho_0^2 } {\ddot {\delta x_k} }
  - { \omega_0^2 \over \rho_0^2 } \sum_{i=1}^{6-p}
 {\partial \over {\partial \theta_i} }
 \bigg ( { {\partial \delta x_k} \over {\partial \theta_i} }
        {\hat g}^{\theta_i \theta_i}  \bigg )
 - {1 \over \lambda_1^2}  \omega_0^2 r_0^2
 \sum_{j=p-1}^{p} {\partial \over {\partial x^j} }
 \bigg ( { {\partial \delta x_k} \over {\partial x^j} }
       \bigg ) = 0 ,~~~(k = 1, \cdots, p-2).
 \label{delxke}
 \ee
 We observe that $\delta x_k$ perturbations decouple from
 ${\delta \rho}$, ${\delta \phi}$ and ${\delta r}$ perturbations.

 Let us introduce the spherical harmonics $Y_l$ on ${\rm S}^{6-p}$,

 \be
 g^{\theta_i \theta_j}
  { \partial \over {\partial \theta_i} }
  { \partial \over {\partial \theta_j} }
  Y_l ( \theta_1 , \dots , \theta_{6-p} ) =
 - Q_l   Y_l ( \theta_1 , \dots , \theta_{6-p} ),
 \ee
 where $Q_l$ is the eigenvalue of the Laplace operator on the unit
 $6-p$ sphere given by

 \be
 Q_l = l(l+ 5-p) , ~~~~~~~ l= 1,2,  \cdots .
 \label{Qlval}
 \ee
 We will not consider the case $l = 0$, which corresponds to zero
 angular momentum on ${\rm S}^{6-p}$.
 Choosing the harmonic oscillation, perturbations can be expressed as

 \bea
 \delta \rho ( t, x_{p-1}, x_p , \theta_1 , \dots , \theta_{6-p} )
 &=& \tilde \delta \rho e^{- i \omega t}
 e^{ i k_{p-1} x_{p-1}}  e^{ i k_p x_p}
 Y_l ( \theta_1 , \dots , \theta_{6-p} ) , \nonumber  \\
 \delta \phi ( t, x_{p-1}, x_p , \theta_1 , \dots , \theta_{6-p} )
 &=& \tilde \delta \phi e^{- i \omega t}
 e^{ i k_{p-1} x_{p-1}}  e^{ i k_p x_p}
 Y_l ( \theta_1 , \dots , \theta_{6-p} ) , \nonumber   \\
 \delta r ( t, x_{p-1}, x_p , \theta_1 , \dots , \theta_{6-p} )
 &=& \tilde \delta r e^{- i \omega t}
 e^{ i k_{p-1} x_{p-1}}  e^{ i k_p x_p}
 Y_l ( \theta_1 , \dots , \theta_{6-p} ) , \nonumber \\
 \delta x_k ( t, x_{p-1}, x_p , \theta_1 , \dots , \theta_{6-p} )
 &=& \tilde \delta x_k e^{- i \omega t}
 e^{ i k_{p-1} x_{p-1}}  e^{ i k_p x_p}
 Y_l ( \theta_1 , \dots , \theta_{6-p} ) , \label{delrprxk}
 \eea
 where $k_{p-1}$ ($k_p$) is the momentum along the $x_{p-1}$
 ($x_p$) direction.
From Eq. (\ref{delxke}), we find the frequency for $\delta x_k$
perturbations as
 \be
 \omega_{x_k}^2 = \omega_0^2 \bigg \{
 Q_l + { {\rho_0^2 r_0^2} \over \lambda_1^2}
 ( k_{p-1}^2 + k_p^2 ) \bigg \} \equiv  \omega_0^2 {Q'}_l ,~~~
 (k = 1, \cdots, p-2).   \label{delxkmode}
 \ee
 The $\delta \rho$, $\delta \phi$ and $\delta r$ perturbations are
 coupled and their normal modes are determined by the
 following matrix equation
 \be
 \pmatrix{
 { 1 \over {1 - \rho_0^2}}
 ( {Q'}_l- {\omega^2 \over \omega_0^2} )&
 i {{5-p} \over \rho_0} {\omega \over \omega_0} &
  {{(5-p)^2} \over 2} {1 \over \rho_0 r_0}  \cr
 - i {{5-p} \over \rho_0} {\omega \over \omega_0} &
 { {1 - \rho_0^2} \over \rho_0^2}
 ( \rho_0^2 {Q'}_l - {\omega^2 \over \omega_0^2} )&
 i {{5-p} \over 2} { {1 - \rho_0^2} \over \rho_0^2 r_0}
 {\omega \over \omega_0}
 \cr
 {{(5-p)^2} \over 2} {1 \over \rho_0^3 r_0} &
 -i {{5-p} \over 2} { {1 - \rho_0^2} \over \rho_0^4 r_0}
 {\omega \over \omega_0}   &
 { 1 \over {\rho_0^2 r_0^2}}
 \bigg \{  {Q'}_l
  + {{5-p} \over 2} ( 4-p - {{5-p} \over 2} {1 \over \rho_0^2} )
   - {\omega^2 \over \omega_0^2} \bigg \}
  \cr }
 \pmatrix{ \tilde \delta \rho \cr
           \tilde \delta \phi \cr
           \tilde \delta r  \cr    }
 ~=~0.     \label{3by3mtx}
 \ee
 Defining $X \equiv \omega / \omega_0 $, the normal modes of the
 coupled equations can be found from the following equation
 obtained from the determinant of the matrix

 \bea
  ( X^2 &-& {Q'}_l ) ( X^2 - \rho_0^2 {Q'}_l )
 \bigg \{ X^2 - {Q'}_l -{n \over 2} ( n-1 - { n \over {2 \rho_0^2} })
 \bigg \}
 + {n^4 \over 4}
 {{1 - \rho_0^2} \over \rho_0^2} (X^2 + \rho_0^2 {Q'}_l )
 \nonumber   \\
  &-& n^2 X^2
  \bigg \{ X^2 - {Q'}_l -{n \over 2} ( n-1 - { n \over {2 \rho_0^2} })
 \bigg \}
  - {n^2 \over 4} { {1-\rho_0^2 } \over \rho_0^2 } X^2
  ( X^2 - {Q'}_l )  = 0 ,  \label{Xequation}
 \eea
 where we defined $n \equiv 5-p$ for simplicity of the expression.

 \subsection{Stability analysis}

 The condition for the giant graviton to be stable over the
 perturbation is that Eq. ({\ref{Xequation}) have all real
 roots. The existence of imaginary part in $\omega$
 means that the $e^{- i \omega t}$ term can grow exponentially, which
 means the instability of the configuration. We will check
 whether it has all real roots or not from the functional analysis.
 For this purpose, we define a function $f(y)$ such that
\bea
 f(y) = && ( y - {Q'}_l ) ( y - \rho_0^2 {Q'}_l )
 \bigg \{ y - {Q'}_l -{n \over 2} ( n-1 - { n \over {2 \rho_0^2} })
 \bigg \}
 + {n^4 \over 4}
 {{1 - \rho_0^2} \over \rho_0^2} (y + \rho_0^2 {Q'}_l )
  \nonumber \\
 && ~~~ - n^2 y
  \bigg \{ y - {Q'}_l -{n \over 2} ( n-1 - { n \over {2 \rho_0^2} })
 \bigg \}
  - {n^2 \over 4} { {1-\rho_0^2 } \over \rho_0^2 } y
  ( y - {Q'}_l )  \nonumber  \\
  \equiv && y^3 + c_2 y^2 + c_1 y + c_0 ,  \label{yequation}
 \eea
 where $y = X^2 = {\omega^2 / \omega_0^2}$
 and $c_i$'s are calculated as

 \bea
  c_0 &=& - \rho_0^2 {Q'}_l^3 -
  {1 \over 2} n ( n-1 - { n \over {2 \rho_0^2} }) \rho_0^2 {Q'}_l^2
  + {1 \over 4} n^4 (1-\rho_0^2) {Q'}_l ,
  \nonumber  \\
  c_1 &=& (1 + 2 \rho_0^2) {Q'}_l^2
  + n( n- {1 \over 2} + {{n-1} \over 2} \rho_0^2 ) {Q'}_l
  + {1 \over 4} n^3 ( n -2 ) ,
  \nonumber  \\
  c_2 &=& - (2 + \rho_0^2) {Q'}_l  - {1 \over 4} n ( 5n -2 ) .
 \label{coeffcs}
 \eea
 The condition for $X= \omega / \omega_0$ to have all real roots is
 equivalent to for $f(y) = 0$ to have all non-negative real roots.
 From the graphical analysis, this is satisfied if the following four
 conditions are satisfied. First of all,
 $df/dy = f' (y) = 3y^2 + 2 c_2 y + c_1 = 0 $ should have real
 solution, i.e.

 \be
 {\rm (i)}~~~~~~~~~~~~~~~~~~~~~~~~~~~~~~~~~~~ c_2^2 -3 c_1 \ge 0.
 \ee
 Let the two real roots of $f' (y) = 3y^2 + 2 c_2 y + c_1 = 0 $ be
 $\alpha$ and $\beta$ ($\alpha < \beta$), then

 \be
 {\rm (ii)}~~~~~~~~~~~~~~~~~~~~~~~~~~~~~~~~~~~~
  f(\alpha) f(\beta) \le 0 ,
 \ee
 and the smaller one ($\alpha$) should be positive

 \be
 {\rm (iii)}~~~~~~~~~~~~~~~~~~~~~~
 \alpha = { {- c_2 - \sqrt{c_2^2 - 3c_1} } \over 3} >
   0.
 \ee
 The vertical-axis intercept $f(y=0)$ should be negative

 \be
 {\rm (iv)}~~~~~~~~~~~~~~~~~~~~~~~~~~~~~~~~~ f( y=0) = c_0 < 0.
 \ee

 First we consider the condition (iv).
 One can easily check that for $n=0 (p=5)$,
 $c_0 = - \rho_0^2 {Q'}_l^3$, thus the condition (iv) is automatically
 satisfied for all regions of $0 < \rho_0 \le 1$ and
${Q'}_l = l (l+n) + ( {\rho_0^2 r_0^2} / \lambda_1^2 )
 ( k_{p-1}^2 + k_p^2 ) > 0$. For $n \ne 0$, the
 condition (iv) is satisfied for
 \be
 \rho_0^2 > { {n^2 ({Q'}_l + n^2) } \over
    {4 {Q'}_l^2 + 2 n (n-1) {Q'}_l + n^4 } },
   ~~~~~~~~ (n \ne 0) ,
 \label{c0constraint}
 \ee
 from the first expression of Eq. (\ref{coeffcs}).

 Secondly, let us examine the conditions (i) and (iii).
 These two are satisfied for any $c_1$ and $c_2$ with

 \be
 c_1 > 0 ,~~~~~~~~~~~~~~~~~~c_2 < 0 .
 \ee
 From Eq. (\ref{coeffcs}), the conditions
  $c_1 >0$ and $c_2 < 0 $ are satisfied for

 \be
 \rho_0^2 > -{ { {Q'}_l^2 + n ( n- {1 \over 2} ) {Q'}_l
   + {3 \over 4} n^3 (n-2) } \over
   { 2 {Q'}_l^2 + {1 \over 2} n ( n- 1) {Q'}_l } } ,
 \label{c1condition}
 \ee

 \be
 \rho_0^2 > -{ {  2 {Q'}_l + {1 \over 4}n ( 5n- 2 ) } \over
    {Q'}_l } .
 \label{c2condition}
 \ee
 For $n =0$ Eqs. (\ref{c1condition}) and (\ref{c2condition})
 are is always satisfied for all values of $ 0 < \rho_0 \le 1$ and
 ${Q'}_l > 0$.
  For $n \ne 0$, for the given range of $\rho_0$ in Eq.
 (\ref{c0constraint}), Eqs. (\ref{c1condition}) and (\ref{c2condition})
 are automatically satisfied for all $ 0 < \rho_0 \le 1$ and
 ${Q'}_l > 0$.

 Finally we consider the condition (ii). We can write down the
 condition (ii) explicitly in terms of $c_i$'s.
 Using $3\alpha^2 + 2 c_2 \alpha + c_1 = 0 $, we can
 write

 \be
 f(\alpha) = {2 \over 3} (c_1 - {1 \over 3} c_2^2 ) \alpha
           + c_0 - { {c_1 c_2} \over 9},
 \ee
 and same expression for $f(\beta)$. Then we have

 \bea
 f(\alpha) f(\beta) &=& \bigg \{ {2 \over 3} (c_1 - {1 \over 3} c_2^2 ) \alpha
           + c_0 - { {c_1 c_2} \over 9} \bigg \}
   \bigg \{ {2 \over 3} (c_1 - {1 \over 3} c_2^2 ) \beta
           + c_0 - { {c_1 c_2} \over 9} \bigg \}  \nonumber  \\
 &=&  {4 \over 9} (c_1 - {1 \over 3} c_2^2 )^2 \alpha \beta
    + {2 \over 3} (c_1 - {1 \over 3} c_2^2 ) ( c_0 - { {c_1 c_2} \over 9} )
    ( \alpha + \beta ) + ( c_0 - { {c_1 c_2} \over 9} )^2 .
 \eea
 Since $\alpha$ and $\beta$ are two real roots of
 $3 y^2 + 2 c_2 y + c_1 = 0 $, substituting

 \be
 \alpha + \beta = - {2 \over 3} c_2 ,
 ~~~ \alpha \beta = {1 \over 3} c_1 ,
 \ee
 the condition (ii) can be written as

 \be
  {4 \over 27} c_1 (c_1 - {1 \over 3} c_2^2 )^2
  + {4 \over 9} c_2 (c_1 - {1 \over 3} c_2^2)( c_0 - { {c_1 c_2} \over 9})
  + ( c_0 - { {c_1 c_2} \over 9} )^2 \le 0.
 \ee
 For the given ranges of $0 < \rho_0 \le 1$ and ${Q'}_l >0$,
  this condition is always satisfied for all possible values of $n
  = 0,1,2,3$ ($p=5,4,3,2$).

 We can summarize the above result as follows.
 For $n=0 (p=5)$, the vibration modes are all real for
 all range of $0 < \rho_0 \le 1$. This means that the
 giant graviton configurations are stable for
 $0 < \rho_0 \le 1$.
 For $n \ne 0 (p \ne 5)$, the vibration modes are all real for
 \be
  { {n^2 ({Q'}_l + n^2) } \over
   {4 {Q'}_l^2 + 2 n (n-1) {Q'}_l + n^4 } } < \rho_0^2 \le 1,
   ~~~~~~~~ (n \ne 0).
 \label{rhocondnne0}
 \ee
 So the giant graviton configurations are stable for
 this range of $\rho_0$.

 \subsection{ Spectrum of the fluctuation modes}

 Now we can find the mode solution under the range of $\rho_0$
 where the stable solution can exist. The solution of the
 coupled ($\rho , \phi , r$) mode equation
 (\ref{Xequation}) gives six frequencies
 $\omega = \pm \omega_1 , \pm \omega_2 , \pm \omega_3$ where

 \bea
 { \omega_1^2 \over \omega_0^2 } &=& ( s_+ + s_- )
                    - {c_2 \over 3} , \nonumber  \\
 { \omega_2^2 \over \omega_0^2 } &=& - {1 \over 2} ( s_+ + s_- )
   - {c_2 \over 3} + { { \sqrt{3} i } \over 2 } ( s_+ - s_- ) , \nonumber  \\
 { \omega_3^2 \over \omega_0^2 } &=& - {1 \over 2} ( s_+ + s_- )
   - {c_2 \over 3} - { { \sqrt{3} i } \over 2 } ( s_+ - s_- ) .
 \eea
 Here $s_\pm$ are defined as

 \be
 s_+ = \{ r + ( q^3 + r^2 )^{1 \over 2} \}^{1 \over 3}, ~~~~
 s_- = \{ r - ( q^3 + r^2 )^{1 \over 2} \}^{1 \over 3},
 \ee
 with

 \be
 q = {1 \over 3} c_1 - {1 \over 9} c_2^2 ,~~~~
 r = {1 \over 3} ( c_1 c_2 - 3 c_0 ) - {1 \over 27} c_2^2 .
 \ee

 There are two special cases where the above expression becomes
 simple. One is the case when the size of the giant graviton is maximum.
  For $\rho_0 = 1$, Eq. (\ref{Xequation}) reduces
 to

 \be
  \{ X^2 - {Q'}_l -{1 \over 4} n (n-2) \}
  \{ X^4 - ( 2 {Q'}_l + n^2) X^2 + {Q'}_l^2 \} = 0 ,
 \ee
 and the solution ($X = \omega / \omega_0$) is given by

 \be
 X^2 = {Q'}_l + {1 \over 4} n (n-2) , ~~~~
 X^2 = {Q'}_l + {1 \over 2} n^2 \pm {1 \over 2}
    \sqrt{ 4 {Q'}_l n^2 + n^4 } .
 \ee
 One can easily check that all values of $X^2$ are real and
 positive using Eq. (\ref{Qlval}). We have shown that the
 giant graviton configuration is stable at the point
 $\rho_0 = 1$ regardless of $n$. Note that $\rho_0 = 1$ case
 corresponds to the maximum value of the angular momentum
 $p_\phi$, which is the realization of the stringy exclusion
 principle.

 The other case we will consider is when $p=5$.
 For $n= 5-p =0$, Eq. (\ref{Xequation}) becomes

 \be
  ( X^2 - {Q'}_l )^2 ( X^2 - \rho_0^2 {Q'}_l ) = 0 ,
 \ee
 and the mode frequencies are

 \be
 \omega^2 =  \omega_0^2 {Q'}_l ~~{\rm (degenerate) , }
 ~~~~ \omega^2 =  \rho_0^2 \omega_0^2 {Q'}_l.
 \label{modepeqfive}
 \ee
 Note that the frequency of degenerate mode is the same as that of
 $\delta x_k$ in Eq. (\ref{delxkmode}). For $p=5$, $\delta \rho$,
 $\delta \phi$ and $\delta r$ perturbations are decoupled. The
 degenerate frequency corresponds to $\delta \rho$ and $\delta r$
 perturbation modes and is independent of the size of brane, while
 the frequancy for $\delta \phi$ depend on $\rho_0$.

 \section{Conclusion}

 We studied the stability analysis and found mode frequencies of
 the giant gravitons in the string
 theory background with NSNS B field. We consider the perturbation
 from the stable configurations generated by D$(p-2)$-D$(p)$ branes
 for $2 \le p \le 5$.
 The vibration modes for $x_k$'s($k=1, \cdots, p-2)$ are decoupled
 and the frequencies are all real.
 The vibration modes for $\rho$ , $\phi$ and
 $r$ are coupled. For $p=5$, they are stable independent of the
 size of the brane. For $p \ne 5$, we calculated the range of the
 size of the brane where they are stable. In the limit when the
 angular momentum on ${\rm S}^{6-p}$ is large, i.e. $l$ is large,
 the condition in Eq. (\ref{rhocondnne0}) becomes
 $0 < \rho_0^2 \le 1$. This means that the giant graviton
 configurations are stable for large angular momentum regardless
 of the size of the branes.
 We would like to emphasize that Eq. (\ref{Lambdae1}) is the crucial
 condition in our calculation. It has been discussed in Ref.
 \cite{dtv0008203} that one can draw the giant graviton picture
 whenever this condition is met.

 In the previous work of the vibration modes of giant gravitons in
 the dilatonic backgrounds \cite{kimmyung}, the perturbation along the
 transverse$(r)$ direction was not considered. Only $\rho$ and
 $\phi$ perturbations were considered and they are coupled.
 Their normal modes are determined by a $2 \times 2$ matrix equation.
 If we turn off the NSNS B field by setting $\varphi = 0$($h_p = 1$),
 Eq. (\ref{dsgen}) reduces to the geometry of the dilatonic
 D$(p)$ brane.
 So with $\delta r = 0$, we have from Eq. (\ref{3by3mtx})

 \be
 \pmatrix{
 { 1 \over {1 - \rho_0^2}}
 ( {Q}_l- {\omega^2 \over \omega_0^2} )&
 i {{5-p} \over \rho_0} {\omega \over \omega_0}  \cr
 - i {{5-p} \over \rho_0} {\omega \over \omega_0} &
 { {1 - \rho_0^2} \over \rho_0^2}
 ( \rho_0^2 {Q}_l - {\omega^2 \over \omega_0^2} ) }
 \pmatrix{ \tilde \delta \rho \cr
           \tilde \delta \phi \cr  }
 ~=~0.     \label{2by2mtx}
 \ee
 This gives the frequencies of two modes
 \be
 { \omega_{\pm}^2 \over  \omega_0^2 } = { 1 \over 2 }
 \bigg [ ( 1 + \rho_0^2 ) Q_l + (5-p)^2
 \pm \sqrt{ (5-p)^4 + 2 (5-p)^2 (1 + \rho_0^2) Q_l
       + Q_l^2 (1 - \rho_0^2)^2 } \bigg ],
 \ee
 which is exactly the same result obtained in Ref. \cite{kimmyung}.
 For $p=5$, we have $\omega_+ = \pm \omega_0 \sqrt{ Q_l }$ and
 $\omega_- = \pm \rho_0 \omega_0 \sqrt{ Q_l }$.
 This result corresponds with the solution in Eq.
 (\ref{modepeqfive}). This implies that our analysis is the
 generalization of the previous work.

 In our perturbation, we considered $\omega_0$ and $r_0$ as
 time independent because the background configuration is valid
 only in the near-horizon region.
 The brane motion in the transverse direction generally induces a
 cosmological evolution of the brane universe, called the mirage
 cosmology \cite{migcos}. The relevance of giant
 gravitons to the mirage cosmology was pointed by Youm
 \cite{youm}. Further studies on this issue is expected in future.

 \section*{Acknowledgment}

 We would like to thank Y. J. Yoon and K. M. Chung for useful discussions.
 This work was supported by Korea Research Foundation Grant
(KRF-2001-015-DP0082).


\begin{references}

\bibitem{myers9910053}
R. Myers, JHEP 9912 (1999) 022, hep-th/9910053.
\bibitem{mst0003075}
J. McGreevy, L. Susskind and N. Toumbas, JHEP 0006 (2000) 008,
hep-th/0003075.
\bibitem{sep}
J. Maldacena and A. Strominger, JHEP 9812 (1998) 005,
hep-th/9804085; A. Jevicki and S.  Ramgoolam, JHEP 9904 (1999)
032, hep-th/9902059; P. Ho, S. Ramgoolam and R. Tatar, Nucl. Phys.
B 573 (2000) 364, hep-th/9907145; S. S. Gubser, Phys. Rev. D 56
(1997) 4984,  hep-th/9704195.
\bibitem{adscft}
J. Maldacena, Adv. Theor. math. Phys. 2 (1998) 231,
hep-th/9711200.
\bibitem{gmt0008015}
M. Grisaru, R. Myers and O. Tafjord, JHEP 0008 (2000) 040,
hep-th/0008015.
\bibitem{0008016}
A. Hashimoto, S. Hirano and N. Itzhaki, JHEP 0008 (2000) 051,
hep-th/0008016.
\bibitem{jlee}
J. Lee, Phys. Rev. D64 (2001) 046012, hep-th/0010191.
\bibitem{mikhailov}
A. Mikhailov, JHEP 0011 (2000) 027, hep-th/0010206.
\bibitem{dtv0008203}
S. R. Das, S. P. Trivedi and S. Vaidya, JHEP 0010 (2000) 037,
hep-th/0008203.
\bibitem{camram} J. M. Camino and A. V. Ramallo, hep-th/0107142.
\bibitem{bmmcp} J. C. Breckenridge, G. Michaud and R. C. Myers,
Phys. Rev. D55 (1997) 6438, hep-th/9611174; M. S. Costa and G.
Papadopoulos, Nucl. Phys. B510 (1998) 217, hep-th/9612204.
\bibitem{djm0009019}
S. R. Das, A. Jevicki and S .D. Mathur, Phys. Rev. D 63 (2001)
024013, hep-th/0009019.
\bibitem{kimmyung} J. Y. Kim and Y. S. Myung, Phys. Lett. B 509
(2001) 157, hep-th/0103001.
\bibitem{migcos}
A. Kehagias and E. Kiritsis, JHEP 9911 (1999) 022, hep-th/9910174;
E. Papantonopoulos and I. Pappa, Mod. Phys. Lett. A15 (2000) 2145,
hep-th/0001183; Phys. Rev. D63 (2001) 103506, hep-th/0010014; J.
Y. Kim, Phys. Rev. D 62 (2000) 065002, hep-th/0004155; Phys. Rev.
D 63 (2000) 045014, hep-th/0009175; D. Youm, Phys. Rev. D63 (2001)
125019, hep-th/0011024.
\bibitem{youm}
D. Youm, Phys. Rev. D63 (2001) 085010, hep-th/0011290.

\end{references}
\end{document}